\documentclass[aps,prl,reprint,amsmath,amssymb,groupaddress,longbibliography,floatfix]{revtex4-2}
\usepackage{graphicx}
\usepackage{hyperref}
\hypersetup{colorlinks,allcolors=blue}

\begin{document}
\title{Quantum Triticality of Bosonic Atomic-Molecular Mixtures with Feshbach Coupling}
\author{Yuan-Hong Chen}
\author{Dong-Chen Zheng}
\author{Renyuan Liao}
\email{ryliao@fjnu.edu.cn}
\affiliation{Fujian Provincial Key Laboratory for Quantum Manipulation and New Energy Materials, College of Physics and Energy, Fujian Normal University, Fuzhou 350117, China}
\affiliation{Fujian Provincial Collaborative Innovation Center for Advanced High-Field Superconducting Materials and Engineering, Fuzhou, 350117, China}
\date{\today}

\begin{abstract}
We develop a functional integral formulation for a homogeneous bosonic atomic-molecular mixture with Feshbach coupling in three-spatial dimensions. Taking phase stability into account, we establish a rich ground-state phase diagram, which features three regions: molecular superfluid (MSF), atomic-molecular superfluid (AMSF), and phase separation (PS). The system can accommodate up to two tricritical points where the three regions meet, with one tricritical point being intrinsic and the other being conditional. Strikingly, we show that the sound velocity vanishes as the AMSF phase touches on the border of phase separation lines. We find that quantum fluctuations correction to the ground-state energy and quantum depletion of the condensates vary nonmonotonically with Feshbach coupling strength as well as molecular percentage. Correlation functions such as pairing amplitudes, density structure factor and spin density structure factor show characteristic behaviors when the system crosses phase transitions. Our work paves the way for future advancement toward understanding salient physics of atom-molecular mixtures.
\end{abstract}

\maketitle

Endowed with a large set of stable states with strong transitions between them and long coherence time, ultracold ensembles of molecules has emerged as a versatile platform for quantum simulation, metrology, quantum information processing and the study of chemical reactions in the ultracold regime~\cite{CARR09,JULI12,BOHN17,SIMO24,TIM24,DAVI24,TIJS24}. In particular, one of the important achievements is recent experimental observation of coherent and collective reactive coupling between Bose-condensed atoms and molecules near a Feshbach resonance~\cite{CHIN21,CHIN23}, which provides fascinating opportunities to explore the long-sought transition from an atomic Bose-Einstein Condensate (BEC) to a molecular BEC. This has stimulated considerable theoretical efforts in exploring reaction dynamics~\cite{MALL22,SINI22,YANG22,MODA24}~and many-body physics~\cite{BELL23,WANG23}~in this system.

Such experimental success also triggers revival interests in studying pairing up bosonic atoms into molecules, the bosonic analog of the crossover from a BEC to a Bardeen-Cooper-Schrieffer (BCS) superfluid in a Fermi gas, which dated back to two decades ago~\cite{LEO04,STOO04,DUIN04,LEE04,LEO08}. These studies and later extensions to confined systems~\cite{SIMO11,SIMO12,KEEL14,ROUS17} have raised the exciting possibilities of an Ising quantum phase transition between distinct molecular and atomic plus molecular condensates. However, the issue of phase stability has been largely overlooked so far~\cite{MUEL08}, which gives puzzles to the result of recent experimental and theoretical simulation~\cite{WANG23}, where an unstable region near resonance region is suggested. This certainly calls for a convincing theoretical explanation.

It is well established that~\cite{HO96,PU98,AO98,TIMM98} for an ordinary BEC mixture with two components occupying different hyperfine states is mechanically stable when the geometrical average of the intraspecies coupling constants is greater than the modulus of the interspecies coupling constant, i.e. $\sqrt{g_{11}g_{22}}>g_{12}$. Otherwise, the system will experience phase separation or collapse. This naturally raises an interesting question: How does the Feshbach coupling, which are responsible for the interconvertion of atoms and molecules, modifies the stability criteria? All these makes examination of phase stability in the ground-state structure of atom-molecule mixtures an urgent and interesting task.

We consider a three-dimensional homogeneous mixture of bosonic atoms and molecules with a Feshbach coupling $\alpha $, described by the following grand canonical Hamiltonian
\begin{align}
&H-\sum_{\sigma =a,m}{\mu _{\sigma}}N_{\sigma}=-\int{d\mathbf{r}\left( \alpha \psi _{a}^{\dagger}\psi _{a}^{\dagger}\psi _m+\mathrm{H.c.} \right)}+\notag \\
&\int{d}\mathbf{r}\sum_{\sigma}{\psi _{\sigma}^{\dagger}}\left( \frac{\hbar ^2\nabla ^2}{2M_{\sigma}}-\mu _{\sigma}+\sum_{\sigma ^{\prime}}{\frac{g_{\sigma \sigma ^{\prime}}}{2}\psi _{\sigma ^{\prime}}^{\dagger}\psi _{\sigma ^{\prime}}} \right) \psi _{\sigma},
\end{align}
where subscript $\sigma =a,m$ denotes the atoms and the diatomic molecules, respectively, $M_a=M$ is the atomic mass, and $M_m=2M$ is the molecular mass. The coefficient $\alpha$, so-called Feshbach coupling, denotes the atom-molecule interconversion strength, and can be assumed to be real positive as one can always make field redefinition for molecular field $\psi_m$. $g_{aa}\equiv g_a$, $g_{mm}\equiv g_m$ and $g_{am}$ characterizes repulsive atom-atom, molecule-molecule and atom-molecule interaction strength, respectively. $\mu_a\equiv\mu$ is the chemical potential for the atomic gas and $\mu_m=2\mu-\nu$ is the chemical potential for the molecular gas, with detuning $\nu $ represents the molecular binding energy, which is experimentally tunable. The chemical potential $\mu $ is introduced to fix the total number of the particles $N=N_a+2N_m\equiv n\mathcal{V}$ with $n$ being the total density and $\mathcal{V}$ being the volume of the system.

\begin{figure}[t]
\includegraphics[width=\linewidth]{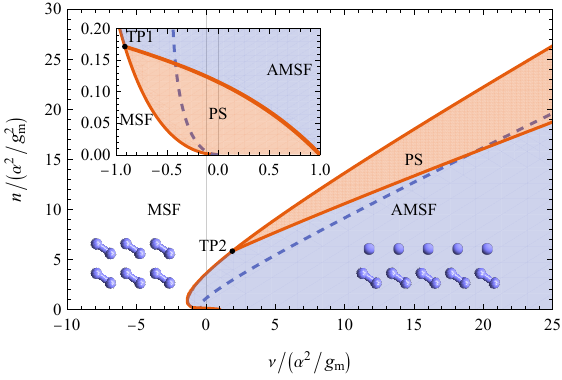}
\caption{Phase diagram spanned by $\nu g_{m}/\alpha ^2 $ and $ng_{m}^{2}/\alpha ^2$ for the interaction ratio $g_{a}:g_{m}:g_{am}=2:1:2$. The phase diagram consists of three regions labeled by molecular superfluid (MSF), atomic and molecular superfluid (AMSF), phase separation (PS). TP1 and TP2 denote two tricritical points. Details near the origin (resonance and low density) are shown in inset.
}
\label{fig1}
\end{figure}

Within the framework of the imaginary-time field integral, the partition function of the system can be cast as $\mathcal{Z} =\int{\mathcal{D} \left[ \psi _{\sigma}^{*},\psi _{\sigma} \right] e^{-S}}$, with the action given by~\cite{SIM23}
\begin{equation}
S=\int_0^{\beta}{d\tau}\left[ H+\sum_{\sigma=a,m}{\int{d\mathbf{r}}\psi _{\sigma}^{*}\left(\partial _{\tau}-\mu_\sigma\right)\psi _{\sigma}} \right],
\end{equation}
where $\beta =1/\left( k_{\mathrm{B}}T \right) $ is the inverse temperature. In the spirit of the Bogoliubov theory, we decompose the Bose field $\psi _{\sigma}$ into a mean-field part $\phi _{\sigma}$ and a fluctuating part $\varphi _{\sigma}$ as $\hat{\psi}_{\sigma}=\phi _{\sigma}+\hat{\varphi}_{\sigma}$ and keep the action up to the quadratic order, yielding an effective action $S_{\mathrm{eff}}=S_0+S_{g}$, where $S_0$ is the mean-field action and $S_g$ is the gaussian action. To facilitate the analysis, we set the condensate wave function for species $\sigma=a,m$ as $\phi _{\sigma}=\sqrt{n_{\sigma}}e^{i\theta _{\sigma}} $ with $\theta_\sigma$ and $n_\sigma$ being the condensate phase and the condensate density, respectively. The Feshbach coupling term $2\alpha \mathrm{Re}\left( \psi _{a}^{2}\psi _{m}^{*} \right) $ clearly locks the atomic and molecular phase together with $\theta _m=2\theta _a\left( \mathrm{mod}\,2\pi \right) $. The mean-field ground energy density $E_{G}^{0}=S_0/\left( \beta \mathcal{V} \right) +\mu n$ becomes
\begin{align}
E_{G}^{0}\left( n_a,n_m \right) ={}&\frac{g_a}{2}n_{a}^{2}+\frac{g_m}{2}n_{m}^{2}+g_{am}n_an_m+\nu n_m\notag \\
&-2\alpha n_a\sqrt{n_m}-\mu \left( n_a+2n_m-n \right) .
\end{align}
The extremum of the ground-state energy requires that
\begin{subequations}
\begin{align}
\frac{\partial E_{G}^{0}}{\partial n_a}={}&g_an_a+g_{am}n_m-2\alpha \sqrt{n_m}-\mu =0,\\
\frac{\partial E_{G}^{0}}{\partial n_m}={}&g_mn_m+g_{am}n_a-\frac{\alpha n_a}{\sqrt{n_m}}-\left( 2\mu -\nu \right) =0.
\end{align}
\end{subequations}
To ensure that it is a minimum, one needs to impose the positive-definite of the hessian matrix $\partial^2E_G^0/\partial n_\sigma \partial n_{\sigma^\prime}$ ($\sigma/\sigma^\prime=a,m$) constructed for the ground-state energy, which yields a density-dependent stability constraint
\begin{equation}
g_a\left( g_m+\frac{\alpha}{\sqrt{n_m}}\frac{n_a}{2n_m} \right) -\left( g_{am}-\frac{\alpha}{\sqrt{n_m}} \right) ^2>0.\label{eq:stability}
\end{equation}
We are now in a position to construct the ground-state phase diagram. To proceed, we define two dimensionless tuning parameters: $C_n\equiv ng_m^2/\alpha^2$ and $C_\nu\equiv\nu g_m/\alpha^2$, which are experimentally relevant and tunable. The resultant ground-state phase diagram spanned by $C_n$ and $C_\nu$ is shown in Fig.~\ref{fig1}. It features three regions: molecular superfluid (MSF), atomic and molecular superfluid (AMSF), and phase separation (PS). The molecular superfluid is characterized by the nonzero value of $\langle\hat{\psi}_m\rangle$ and the vanishing of $\langle\hat{\psi}_a\rangle$, while in the AMSF phase both $\langle\hat{\psi}_m\rangle$ and $\langle\hat{\psi}_a\rangle$ achieves a nonzero value. The phase boundary line defined by $C_{\nu}=\left( g_{am}/g_m-1/2 \right) C_n-2\sqrt{2C_n}$ separates the MSF phase from possible phases of superfluid mixtures. The stability constraint given by Eq.~\eqref{eq:stability} further divides the remaining phase diagram into stable miscible AMSF phase and unstable region toward phase separation. When phase separation occurs, the system will minimize its total ground-state energy through a first-order phase transition by forming spatially distinct homogeneous MSF and AMSF, or two spatially homogenous AMSF, which could be determined by Maxwell construction in principle~\cite{REI16}.

The intersection of the phase boundary line and the phase separation line determines the tricritical point (TP), whose coordinate $\left( C_{\nu}^{TP},C_{n}^{TP} \right) $ affords an analytical expression
\begin{subequations}
\begin{align}
C_{n}^{TP}={}&2C_{\pm}^{2},\label{eq:TP}\\
C_{\nu}^{TP}={}&\left( \frac{g_{am}}{g_m}-\frac{1}{2} \right) C_{n}^{TP}-4C_{\pm},
\end{align}
\end{subequations}
where we have used a shorthand notation $C_{\pm}=1/\left( g_{am}/g_m\pm \sqrt{g_a/g_m} \right) $. The upper sign corresponds to the inherent tricritical point (TP1) that always exists near resonance, while the lower sign corresponds to a conditional tricritical point (TP2) which exists if $g_{am}>\sqrt{g_ag_m}$. Remarkably, this indicates that a finite Feshbach coupling extend the stability of an ordinary two-component BEC mixture. In Fig.~\ref{fig1}, we have chosen the parameters to be $g_a:g_{m}:g_{am}=2:1:2$, which indeed allows the system to accommodate a second tricritical point TP2. To investigate how the tricritical points evolve with interaction parameters, it is convenient to define a dimensionless parameter characterizing the interactions $\gamma \equiv \left( g_{am}-\sqrt{g_ag_m} \right) /g_m$. We consider the case $\gamma>0$ where the system accommodates two tricritical points, then it is straightforward to find from Eq.~\eqref{eq:TP} that $C_n^{TP1}C_n^{TP2}=4/\gamma^4$. On one side when $\gamma$ increases, the tricritical point TP2 run away from TP1 and move toward the infinity, on the other side when $\gamma$ approaches zero, the tricritical point TP2 merges with TP1.

\begin{figure}[t]
\includegraphics[width=\linewidth]{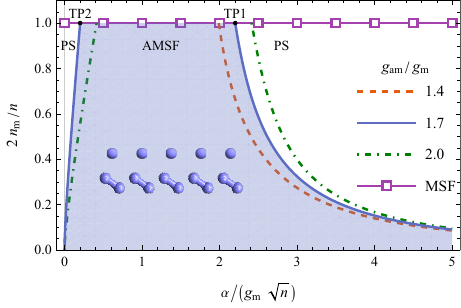}
\caption{%
Phase diagram spanned by the molecular percentage $P\equiv 2n_m/n$ and the dimensionless Feshbach coupling $\alpha /\left( g_m\sqrt{n} \right) $ for three different atom-molecular interaction parameters $g_{am}/g_m=1.4$, $1.7$ and $2.0$ while $g_a/g_m=2$ is kept fixed. The phase diagram consists of three regions labeled by molecular superfluid (MSF), atomic and molecular superfluid (AMSF), phase separation (PS). For $g_{am}/g_m=1.7$ and $2.0$, the system hosts two tricritical point intersecting the three regions, while for $g_{am}/g_m=1.4$ the system hosts only one tricritical point.
}
\label{fig2}
\end{figure}

To reveal the fascinating effects of the molecular percentage and Feshbach coupling in shaping our system, we construct the ground-state phase diagram spanned by molecular percentage $P\equiv2n_m/n$ and dimensionless Feshbach coupling $\tilde{\alpha}\equiv \alpha /\left( g_m\sqrt{n} \right) $ for three typical atomic-molecular interaction strengths $g_{am}/g_m=1.4$, $1.7$, and $2.0$, as shown in Fig.~\ref{fig2}. The phase diagram features three distinct regions: the molecular superfluid (MSF), the atomic and molecular superfluid (AMSF) and phase separation (PS). By definition, the MSF phase is exclusively located on the horizontal line of $P=1$, which also serves as the phase boundary line separating the MSF phase and the AMSF phase. The phase stability constraint determines the phase separation line, which is given by
\begin{equation}
\frac{g_a}{g_m}\left( 1+\sqrt{\frac{2}{P}}\frac{1-P}{P}\tilde{\alpha} \right) =\left( \frac{g_{am}}{g_m}-\sqrt{\frac{2}{P}}\tilde{\alpha} \right) .\label{eq:psline}
\end{equation}
When $\gamma>0$, which are the cases for $g_{am}/g_m=1.7$ and $2.0$, Eq.~\eqref{eq:psline} hosts two phase separation lines. The left line determines a lower bound for a critical Feshbach coupling $\tilde{\alpha}$ at a given molecular percentage $P$ to sustain the AMSF phase, while the right line gives an upper bound for a critical Feshbach coupling $\tilde{\alpha}$. In contrast, when $\gamma<0$, which is the case for $g_{am}/g_m=1.4$, Eq.~\eqref{eq:psline} hosts only the right phase separation line, determining an upper bound for a critical Feshbach coupling at a given molecular percentage $P$ to sustain the AMSF phase.

We proceed to consider the effects of quantum fluctuation on the properties of the system residing within the AMSF phase. The gaussian action on top of the mean-field configuration is given by $S_{g}=\int{d\tau d\mathbf{r}}\mathcal{L} _{g}$ with
\begin{align}
\mathcal{L} _g={}&\left[ \left( a\varphi _{a}^{*}\varphi _m+\frac{b}{2}\varphi _{a}^{2}+c\varphi _a\varphi _m+\frac{d}{2}\varphi _{m}^{2} \right) +\mathrm{c.c} \right] \notag \\
&+\sum_{\sigma =a,m}{\varphi _{\sigma}^{*}}\left( \partial _{\tau}+\hat{\epsilon}_{\sigma} \right) \varphi _{\sigma},
\end{align}
where we have adopted the shorthand notations as $\hat{\epsilon}_\sigma=-\hbar^2\nabla^2/2M_\sigma-\mu_\sigma+2g_\sigma n_\sigma+g_{am}n_{\bar{\sigma}}$, $a=g_{am}\sqrt{n_{a}n_{m}}-2\alpha \sqrt{n_{a}}$, $b=g_{a}n_{a}-2\alpha \sqrt{n_{m}}$, $c=g_{am}\sqrt{n_an_{m}}$ and $d=g_mn_m$. To exploit the translational invariance of the system, we make Fourier transformation $\varphi_\sigma=\sum_q\varphi_{qa}\exp{(iqx)}$ where $q=(\mathbf{q},\omega_n)$ and $x=(\mathbf{r},\tau)$ are 4-vectors. To represent the gaussian action in a concise form, we shall define a column vector $\varPhi_q=(\varphi_{qa}\varphi_{qm}\varphi_{-qa}^\dagger\varphi_{-qm}^\dagger)^T$. Then the gaussian action is written compactly as $S_g=\sum_q\varPhi_q^\dagger\mathcal{G}^{-1}\varPhi_q/2-\beta\sum_\mathbf{q}\left(\epsilon_{\mathbf{q}a}+\epsilon_{\mathbf{q}m}\right)/2$, where $\epsilon_{\mathbf{q}a}=q^2/2M+g_an_a+2\alpha\sqrt{n_m}$, $\epsilon_{\mathbf{q}m}=q^2/4M+g_mn_m+\alpha n_a/\sqrt{n_m}$, and the inverse Green's function $\mathcal{G}^{-1}(\mathbf{q},z)$ is defined as follows
\begin{equation}
\mathcal{G} ^{-1}=\begin{pmatrix}
-z+\epsilon _{\mathbf{q}a}& a& b& c\\a& -z+\epsilon _{\mathbf{q}m}& c& d\\
b& c& z+\epsilon _{\mathbf{q}}& a\\c& d& a& z+\epsilon _{\mathbf{q}m}\end{pmatrix}.\label{eq:GreenFun}
\end{equation}
From now on, we shall adopt natural units by setting $\hbar =2M=k_{\mathrm{B}}=1$ for brevity. We choose $g_mn$ as the basic energy scale, and the corresponding momentum scale is $\sqrt{g_mn}$.

The excitation spectrum corresponds to the poles of the Green's function and could be found by seeking solutions of the secular equation $\det \mathcal{G} ^{-1}\left( \mathbf{q},\omega \right) =0$. It accommodates two branches of excitation spectrum
\begin{equation}
\omega _{\pm}\left( \mathbf{q} \right) =\sqrt{A_{\mathbf{q}}\pm \sqrt{B_{\mathbf{q}}^{2}+C_{1\mathbf{q}}C_{2\mathbf{q}}}}, \label{eq:excitation}
\end{equation}
where we have defined the relevant parameters: $A_{\mathbf{q}}=\left( \epsilon _{\mathbf{q}a}^{2}+\epsilon _{\mathbf{q}m}^{2}-b^2-d^2 \right) /2+a^2-c^2$, $B_{\mathbf{q}}=\big( \epsilon _{\mathbf{q}a}^{2}-\epsilon _{\mathbf{q}m}^{2}-b^2+d^2 \big) /2$, $C_{1\mathbf{q}}=\left( a-c \right) \left( \epsilon _{\mathbf{q}a}+b \right) +\left( a+c \right) \left( \epsilon _{\mathbf{q}m}-d \right) $ and $C_{2\mathbf{q}}=\left( a+c \right) \left( \epsilon _{\mathbf{q}a}-b \right) +\left( a-c \right) \left( \epsilon _{\mathbf{q}m}+d \right) $. In the absence of Feshbach coupling $\alpha=0$, the excitation spectrum Eq.~\eqref{eq:excitation} reduces to two gapless excitations corresponding to two Goldstone modes of the homogenous binary BEC mixture~\cite{AO00}. A finite Feshbach coupling explicitly breaks $U(1)\times U(1)$ symmetry of the $\alpha=0$ Hamiltonian down to $U(1)\times \mathbb{Z}_2$, since the Hamiltonian is now invariant under the transformation $\psi_a\rightarrow e^{i\theta}\psi_a$ and $\psi_m\rightarrow e^{i2\theta}\psi_m$. This $U(1)\times \mathbb{Z}_2$ symmetry is completely broken in the AMSF phase. In consequence, the low-energy excitations of the system features one gapless mode corresponding to in-phase fluctuations of the phases $\theta_a$ and $\theta_m$ of the atomic and molecular condensates, and one gapped mode associated with the out-of-phase fluctuations $2\theta_a-\theta_m$ of the atomic and molecular condensates.

\begin{figure}[t]
\includegraphics[width=\linewidth]{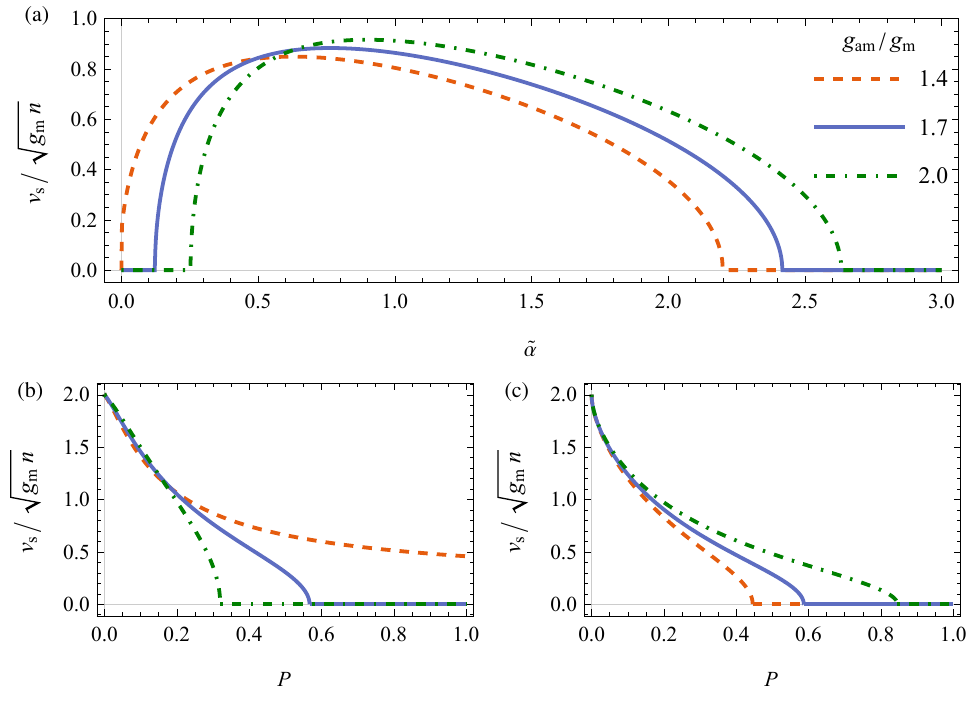}
\caption{The sound velocity $v_s/ \sqrt{g_mn}$ associated with the gapless branch $v_s=\lim_{q\rightarrow0}\omega_-(\mathbf{q})/q$ for three typical atomic-molecular interaction strengths $g_{am}/g_m=1.4$, $1.7$ and $2.0$. (a) As a function of dimensionless Feshbach coupling strength $\tilde{\alpha}\equiv \alpha /\left( g_m\sqrt{n} \right) $ at $n_a=n_m$. (b), (c) As a function of molecular percentage $P\equiv2n_m/n$ at $\tilde{\alpha}=0.1$ and $2.5$, respectively. Here we have set $g_a/g_m=2.0$.
}
\label{fig3}
\end{figure}

The behaviors of the sound velocity associated with the gapless branch $v_s=\lim_{q\rightarrow 0} \omega _-\left( \mathbf{q} \right) /q$ for three typical atomic-molecular coupling strengths $g_{am}/g_m=1.4$, $1.7$ and $2.0$ are shown in Fig.~\ref{fig3}. Evidently, the sound velocity shows a nonmonotonic trend with varying Feshbach coupling $\tilde{\alpha}$, as in panel (a). At small Feshbach coupling, the sound velocity increases rapidly with $\tilde{\alpha}$, then it develops a maximum at some optimal Feshbach coupling, and finally it decreases and drops to zero with increasing $\tilde{\alpha}$. For $g_{am}/g_a=1.7$ and $2.0$, where the system has two phase separation lines, the sound velocity drops to zero at two critical Feshbach coupling. While for $g_{am}/g_m=1.4$ where the system has only one phase separation line, the sound velocity drops to zero at one critical Feshbach coupling. The behaviors of the sound velocity as a function of the molecular percentage $P$ at fixed Feshbach coupling are shown in panels (b) and (c). For a small Feshbach coupling $\tilde{\alpha}=0.1$ shown in panel (b) , in cases ($g_a/g_m=1.7$ and $2.0$) where the system has a left separation line, there exists a threshold of $P$ above which the sound velocity vanishes. While for $g_a/g_m=1.4$, the sound velocity stays finite up to $P=1$. For a large Feshbach coupling $\tilde{\alpha}=2.5$ shown in panel (c), the sound velocity decreases with increasing $P$ until it drops to zero at some threshold of $P$, above which the system is unstable toward phase separation. Therefore, the vanishing of the sound velocity at a critical Feshbach coupling or a critical molecular percentage unambiguously signals that the system is unstable toward phase separation.

\begin{figure}[t]
\includegraphics[width=\linewidth]{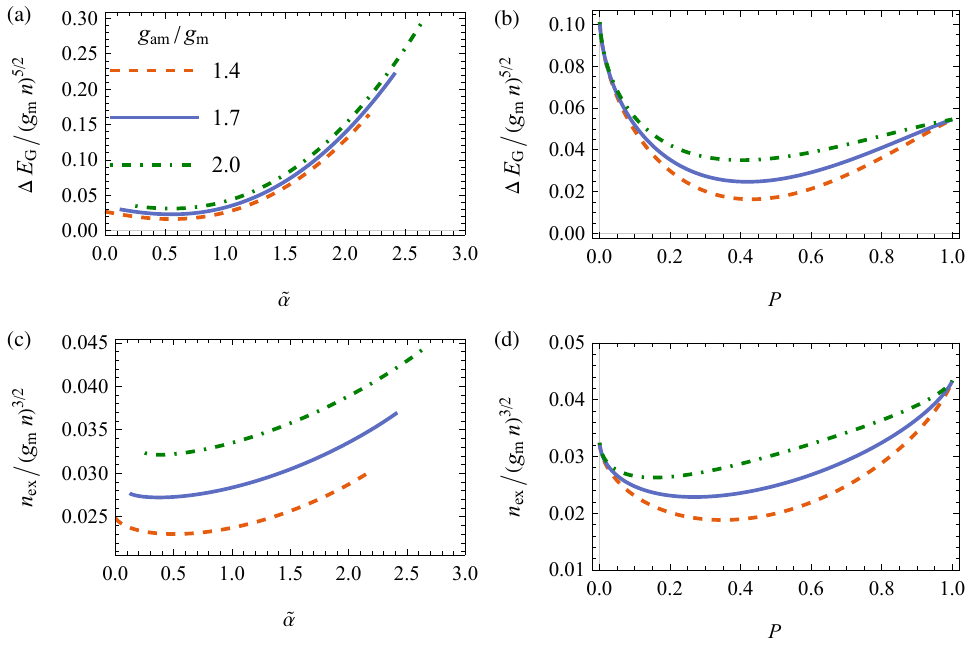}
\caption{Quantum fluctuation correction to the ground-state energy density and quantum depletion of condensates for three typical atomic-molecular interaction strengths $g_{am}/g_m=1.4$, $1.7$ and $2.0$. Shown in the upper panels are quantum fluctuation correction to the ground-state energy density $\Delta E_G$. (a) As a function of dimensionless Feshbach coupling $\tilde{\alpha}$ with $n_a=n_m$. (b) As a function of molecular percentage $P$ with $\tilde{\alpha}=1.0$. Shown in the lower panels are the number density of excited particles $n_{ex}$ due to quantum fluctuation. (c) As a function of dimensionless Feshbach coupling $\tilde{\alpha}$ with $n_a=n_m$. (d) As a function of molecular percentage $P$ with $\tilde{\alpha}=1.0$. Here we have set $g_a/g_m=2.0$.
}
\label{fig4}
\end{figure}

The thermodynamic potential is given by $\Omega =-\ln \mathcal{Z} /\beta =S_0/\beta +\mathrm{Tr}\,\ln \mathcal{G} ^{-1}/\left( 2\beta \right) -\sum_{\mathbf{q}}{\left( \epsilon _{\mathbf{q}a}+\epsilon _{\mathbf{q}m} \right) /2}$.
It should be pointed out that the interaction parameters $g_a$, $g_m$ and $g_{am}$ as well as $\alpha$ and $\nu$ contained in $S_0$ needs to be renormalized~\cite{AND04,STO09} to ensure nondivergent behaviors. Correspondingly, the quantum correction to the ground-state energy density $\Delta E_G$ can be renormalized by subtracting counter-terms (for details, please see Supplemental Material~\cite{SM24})
\begin{align}
\Delta E_G={}&\frac{1}{\mathcal{V}}\sum_{\mathbf{q}}{\left( \frac{\omega _+(\mathbf{q})+\omega _-(\mathbf{q})-\epsilon _{\mathbf{q}a}-\epsilon _{\mathbf{q}m}}{2} \right)}\notag \\
+{}&\frac{1}{\mathcal{V}}\sum_{\mathbf{q}}{\left( \frac{b^2/2}{q^2/2m_1}+\frac{d^2/2}{q^2/2m_2}+\frac{c^2}{q^2/2m_3} \right)},
\end{align}
where in the above $m_1=M/2$, $m_2=M$, $m_3=2M/3$ is the effective mass for atom-atom, molecule-molecule and atom-molecule scattering process, respectively. It is easy to show that such way of regularization is consistent with previous schemes~\cite{WANG23,HOLL02,ZHAI21}. We show the quantum correction to the ground-state energy in the upper panels of Fig.~\ref{fig4}. It is interesting to notice that $\Delta E_G$ decreases slowly with $\tilde{\alpha}$ when $\tilde{\alpha}$ is small and the trend is reversed when $\tilde{\alpha}$ is large. This suggests that there exists an optimal $\tilde{\alpha}$ under which the quantum fluctuation correction to the energy achieves a minimum. The dependence of $\Delta E_G$ on the molecular percentage $P$ is shown in panel (b). It decreases rapidly with $P$ when $P$ is small until it reaches a minimum, then it increase slowly with the molecular percentage.

For the system to be stable, we require that the quantum depletion of the condensates should be finite. We evaluate the density of the exited particles for species $\sigma$ due to quantum fluctuation via Green's function $n_{ex,\sigma}=\sum_{\mathbf{q},i\omega _n}{\mathcal{G} _{\sigma \sigma}\left( \mathbf{q},i\omega _n \right)}$. By analyzing the low-energy asymptotic behavior of the excitation spectrum, we have verified that there is no infradivergence, which renders that $n_{ex}=n_{ex,a}+2n_{ex,m}$ is a finite quantity. We show the depletion of the condensates $n_{ex}$ in the lower panels of Fig.~\ref{fig4}. As indicated in panel (c), for a fixed $P$, there exists an optimal $\tilde{\alpha}$ under which the quantum depletion of the condensates develops a minimum. Correspondingly, for a fixed $\tilde{\alpha}$, there exits an optimal molecular percentage under which $n_{ex}$ achieves a minimum. As expected, a higher interspecies coupling $g_{am}$ leads to a larger quantum depletion.

\begin{figure}[t]
\includegraphics[width=\linewidth]{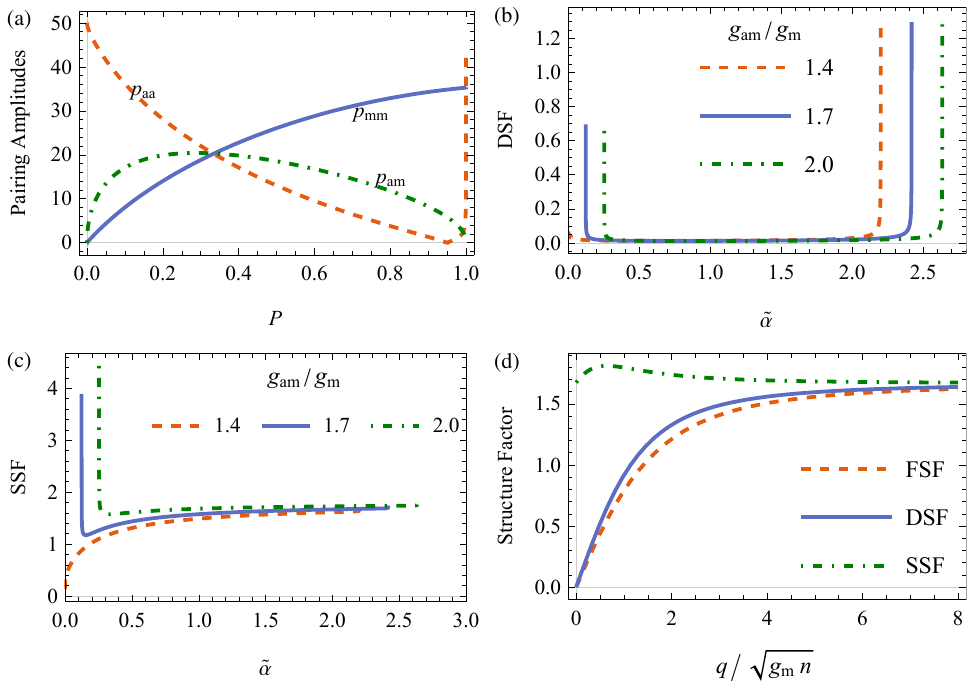}
\caption{Correlation functions at $g_a/g_m=2.0$ for three typical atomic-molecular interaction strengths $g_{am}/g_m=1.4$, $1.7$ and $2.0$. (a) Pairing amplitudes ($p_{aa}$, $p_{mm}$ and $p_{am}$) as a function of the molecular percentage $P$. (b) Density structure factor (DSF) and (c) Spin density structure factor (SSF) as a function of dimensionless Feshbach coupling $\tilde{\alpha}$ at $n_a=n_m$. (d) Momentum dependence of DSF and SSF at $n_a=n_m$. For comparison, we have constructed the Feynman density structure factor (FSF) from Feynman relation. In panel (a), (b) and (c), we have fixed the momentum $q/\sqrt{g_mn}=0.01$. In panel (a) and (d), we have fixed the dimensionless Feshbach coupling $\tilde{\alpha}=1.0$.
}
\label{fig5}
\end{figure}

Correlation functions give important signatures when the system crosses a phase transition, as illustrated in Fig.~\ref{fig5}. The pairing amplitudes $p_{\sigma \sigma ^{\prime}}(\mathbf{q})\equiv |\langle \varphi _{\mathbf{q}\sigma}\varphi _{-\mathbf{q}\sigma}\rangle |$ are shown in panel (a) with varying molecular percentage $P$. As expected, the atom-atom pairing amplitude $p_{aa}$ decreases steadily as the molecular percentage increases from zero, while the trend of molecule-molecule pairing amplitude is reversed. As a comprise, the atom-molecule pairing amplitude shows a nonmonotonic behavior. Strikingly, when the molecular percentage approaches unity, $p_{aa}$ increases from zero vary rapidly. Such revival of pairing correlation signals the buildup of atom pairs when the system crosses the phase transition between AMSF and MSF.

The density (spin density) structure factor $S^{d\left( s \right)}\left( \mathbf{q} \right) \equiv \langle\delta \rho _{\mathbf{q}}^{d\left( s \right) \dagger}\delta \rho _{\mathbf{q}}^{d\left( s \right)}\rangle/N$ probes the density (spin density) fluctuations of the system, where $\rho _{\mathbf{q}}^{d}=\rho _{\mathbf{q}}^{a}+2\rho _{\mathbf{q}}^{m}$ and $\rho _{\mathbf{q}}^{s}=\rho _{\mathbf{q}}^{a}-2\rho _{\mathbf{q}}^{m}$ is the density and spin density, respectively. Shown in panel (b), the density structure factor (DSF) shows divergent behavior at a small momentum when the system crosses both phase separation lines. In stark contrast, the spin structure factor (SSF) shows divergent behavior only when the system crosses the left phase separation line characterized by a small Feshbach coupling. The momentum dependences of DSF and SSF are shown in panel (d) with $\tilde{\alpha}=1$ and $g_a/g_m=2.0$. While the spin density structure factor is insensitive to the variation of the momentum, the density structure factor DSF increases steadily with momentum magnitude and saturates at large momenta. Interestingly, at small and large momenta, apart from a scaling factor, Feynman relation~\cite{FEYN54} connecting static structure factor and low-energy excitation spectrum is largely satisfied. For comparison, we have constructed Feynman structure factor as FSF $\equiv \left( 1+P \right) \left( q^2/2 \right) /\omega _-$. Clearly, at a small momentum, $S(\mathbf{q})$ probes the collective excitations while at a large momentum it probes the single-particle excitation of molecular condensates. The density structure factor is routinely measured with the Bragg spectroscopy~\cite{STAM99}. The spin structure factor can be measured from noise correlations or Bragg scattering of light~\cite{HART15,MAZU17}.

In summary, we have conducted comprehensive studies on the ground state properties of bosonic atomic-molecular mixtures with a Feshbach coupling. We have found rich phase diagrams characterized by up to two tricritical points. We characterized the AMSF phase and associated phase transitions with various physical quantities. Our work lays down a solid framework to explore salient physics of atom-molecular mixtures at finite temperatures~\cite{STRI19,YU20}. We expect that experimental verification of our predictions will add new vitality to the flourishing development of ultracold atomic and molecular gases.

This work is supported by NSFC under Grants No.12174055 and No.11674058, and by the Natural Science
Foundation of Fujian under Grant No. 2020J01195.

\bibliography{MolecularBEC}
\end{document}